\documentclass{fcs}
\usepackage{bm}
\usepackage{longtable}
\volumn{ }
\doi {: 10.1093/comnet/xxx000 }
\articletype{REVIEW~ARTICLE}

\copynote{{\copyright} Higher Education Press and Springer-Verlag Berlin Heidelberg 2022}
\ratime{Received August 23, 2022} 
\email{$ nasim.abdullah@ieee.org $}
\title{$\bm{A~ Survey~ of~ Recommender~ System~ Techniques~ and~ the~ E-commerce~ Domain}$}
\author{Imran Hossain , Md Aminul Haque Palash , Anika Tabassum Sejuty , Noor A Tanjim , MD Abdullah AL Nasim \xff , Sarwar Saif , Abu Bokor Suraj, Md Mahim Anjum Haque, Nazmul Karim }
\address{
{\quad Department of Research and Development, Pioneer Alpha, Bangladesh}}
\begin{document}
\maketitle
\setcounter{page}{1}
\setlength{\baselineskip}{14pt}

%\begin{document}
%\maketitle
%\setcounter{page}{1}
%\setlength{\baselineskip}{14pt}

\begin{abstract}
In this big data era, it is hard for the current generation to find the right data from the huge amount of data contained within online platforms. In such a situation, there is a need for an information filtering system that might help them find the information they are looking for. In recent years, a research field has emerged known as recommender systems. Recommenders have become important as they have many real-life applications. This paper reviews the different techniques and developments of recommender systems in e-commerce, e-tourism, e-resources, e-government, e-learning, and e-library. By analyzing recent work on this topic, we will be able to provide a detailed overview of current developments and identify existing difficulties in recommendation systems. The final results give practitioners and researchers the necessary guidance and insights into the recommendation system and its application.
\end{abstract}

\Keywords{Recommendation Techniques , Recommender System , Collaborative Filtering, Scalability , E-commerce.}

\section{Introduction}
\noindent Recommendation systems are software devices and techniques that make recommendations to users for suitable and interesting products. The recommendations apply to different processes of decision-making, such as which items to purchase, which news to read, or which music to listen to. The recommendation has become more and more relevant and has changed the way of interaction between users and websites. Recommendation engines are developed to recommend songs, products, films, television shows, and a variety of other content that involves some degree of choice. The fundamental function is to define users' needs and suggest them based on what other users have looked at, are interested in, and enjoyed. Recommendation systems have a wide range of applications in several fields, such as economics, e-commerce, education, and scientific research papers \cite{yu2018pave}. 

The expeditious development of IT is rapidly increasing the amount of data. Recommender systems use information processing to suggest user-interesting information, which can be defined as the process that recommends a suitable product after knowing the expectations and wishes of the customer \cite{kotsogiannis2017directed}. Most of the literature on recommendation systems focused on improving recommendation technique's reliability. They usually contain input data, background data, and an algorithm that combines input data and background data to provide a recommendation. Therefore, it saves time and unnecessary effort in finding products, and they will recommend the right products to purchase \cite{florea2017spark}. Such systems are also particularly useful for recognizing the user's preference. 

Recommendation systems help users navigate through website data by making suggestions based on their preferences \cite{komkhao2013incremental}. Developing recommender systems should be an effective solution to help customers quickly pick items based on their interests. To provide the user with a more effective search environment \cite{jiang2015author}, they combine concepts of machine learning, information retrieval, and user recognition. Such systems are designed to determine user interests, behaviors, and help them get what they're searching for or find useful in a large amount of data. 

In all e-commerce, e-tourism, e-government, e-government software, and information management systems, recommender systems are mostly used. Recommender systems are responsible for providing specific users with accurate and reliable data. It is a serious business system that reshapes the e-environment world. Most major e-commerce companies employ recommendation systems to help their customers find items they may be interested in buying quickly, leading to easier checkouts and more sales. Essentially, a recommender program is an adaptive framework based on user experience and user data analytics to define and serve the needs of the user. There are collaborative and content-based, knowledge-based methods to carry out such an analysis. This artificial intelligence explores huge amounts of user data, including clicks, orders, ratings, etc., using various algorithms to make recommendations \cite{mustafa2015performance}. 

Many research articles have been written over the past few years on recommender systems. Such papers, however, concentrate on recommendation techniques and methods or a particular area of development of recommendation systems. Only the comprehensive study of recommendation systems applications is discussed in the paper \cite{lu2015recommender}. Since recommendation systems are playing an important role in many areas, this paper aims at analyzing the current situation of recommendation systems and discussing various forms of recommendation techniques and system development in real-life software applications. 

Two types of articles have been reviewed and classified in this research: 
\begin{enumerate}
  \item Articles based on recommender system techniques and related issues.
  \item Articles based on applications of recommender systems.
\end{enumerate} 

The selection process is based on the following criteria for choosing high-impact and creative articles: 1) published recent year 2) published in journals and conference proceedings which has a good impact factor. 
Such papers are reviewed and chosen as follows: 
\begin{enumerate}
 \item For selecting research papers on RSs for our review, the following journal repositories were used: ACM Digital Library, IEEE, Elsevier, and Springer.
 \item Searching process was therefore carried out based on keywords  relevant to RSs in e-commerce, e-learning, and e-library,   e-tourism, e-resources, e-government applications, etc.
 \item Depending on the keywords, recommender system techniques such as content-based filtering, collaborative filtering, hybrid approaches, context-aware, and knowledge-based strategies separated these articles. Relevant papers were examined and chosen for each RS method.
\end{enumerate}
Finally, we selected 75 papers for our research on RS. 
The following section of this paper is described as Section 2 introduces and categorizes the recommendation system techniques. Sections 3–7 frequently discuss recommendation systems in e-commerce, e-learning and e-library, e-tourism, e-government, and e-resources domains. Section 8 discusses the challenge of recommendation systems, and Section 9 provides comprehensive analysis. 

\section{Recommendation techniques}
\begin{figure}[!ht]
\centering
\includegraphics[scale=0.3]{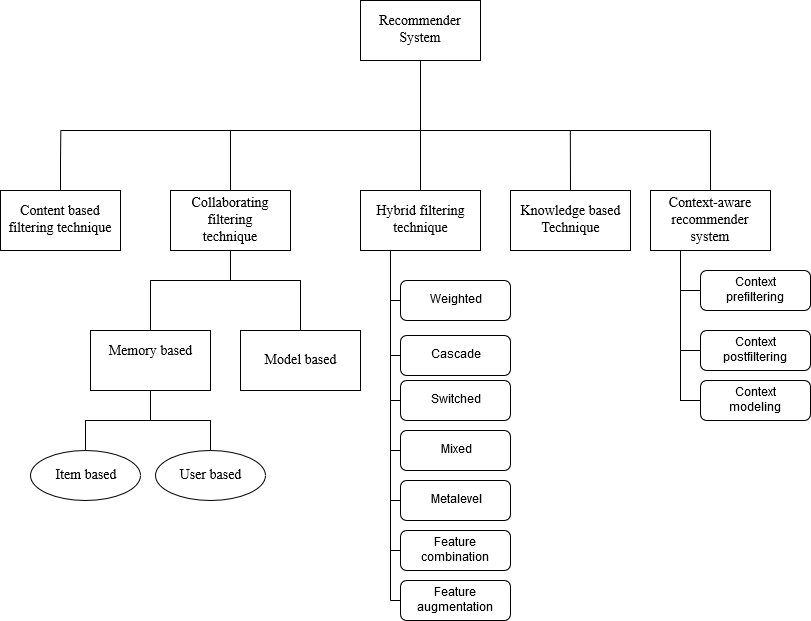}
\caption{Classification of Recommender System}
\label{Classification of Recommender System}
\end{figure}

Using effective and appropriate recommendation systems is essential for a platform that will provide its users with helpful and informative recommendations. We divided the recommender system into i) content-based filtering, ii) collaborative filtering, iii) hybrid filtering, iv) context-awareness, and v) knowledge-based recommender techniques. Fig. \ref{Classification of Recommender System} displays the classification of recommendation techniques. 
 
\subsection{Content Based Filtering Technique} 
In this subsection, we discuss the content-based filtering technique. 

The content-based system makes recommendations for those products that are close to the products already used by customers in the past \cite{burke2002hybrid}. To generate suggestions, the CB method conducts more assessments of the item's attributes. In web pages, journals, product, and news recommendations, the CBF filtering method is most effective.

The content-based system is an algorithm that depends on the domain to produce predictions. It focuses more on analyzing the characteristics of products. Using characteristics obtained from the content of products that the user has assessed in the previous, recommendations are produced in content-based filtering techniques oriented on customer profiles. The user has advised items mostly linked to the favorably rated items. To produce significant suggestions, CBF utilizes distinct kinds of models to discover similarities between records. It does not take into account the attitude of customers towards the product, which limits its recommendation accuracy \cite{li2015accurate}.

Recommendations based on content treat recommendations as a user-specific classification issue and train a classifier based on product characteristics for the user's likes and dislikes. Specifically, based on the characteristics of the content, similarity metrics can identify the most similar items that will be rated similarly \cite{bauer2014recommender}. It also proposes products similar to those that a particular customer enjoyed in the list of item rates. 

\begin{figure}[!ht]
 	\centering
    \includegraphics[scale=0.4]{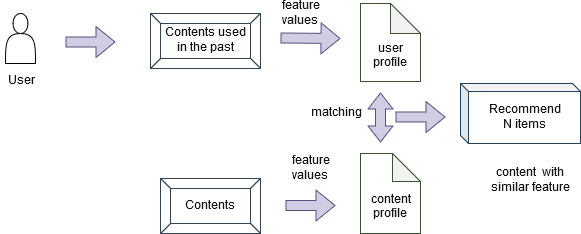}
    \caption{Content based filtering process}
 	\label{Content based filtering process}
\end{figure}

The goals of content-based methods depend on discovering correlations between item content and user correlation, as with CF methods. The root of the strategy based on content can be directly linked to the retrieval of information \cite{champiri2015systematic}. Fig. \ref{Content based filtering process} shows the process of CB filtering.
\break
A commonly used algorithm is the representation of tf–idf. By using the weighted vector of item characteristics, the scheme generates a content-based user profile. The profile of the customer is built by evaluating user answers to user navigation history. It indicates products with similar characteristics to those chosen by the user in the past \cite{pazzani2007content}. Simple methods use overall rated item vector values, but other advanced schemes use machine learning techniques such as Clustering, Bayesian Network, Artificial Neural Network, Decision Tree, or Neural Network to predict the likelihood of the item \cite{zhu2009personalized}. In this paper \cite{kim2011recommender}  author submitted a customized model based on a probabilistic relational model. Furthermore, the CBF approach can still update its recommendation in a short time if the user profile gets changed. 

The main drawback of this method is the need for a deep understanding and description of the characteristics of the products in the profile. However, by pairing CB with the CF filtering strategy and other methods, these shortcomings can be resolved.
\subsection{ Collaborative filtering technique} 
\label{subsec1}
In this subsection, we discuss the collaborative filtering technique.

Collaborative recommendations are most familiar and commonly applied, particularly in e-commerce \cite{burke2002hybrid} and one of the most effective methods used in the recommendation \cite{kuzelewska2011advantages}. Collaborative filtering techniques generate user-specific suggestions for products based on rating or buying items without requiring explicit data on either products or users. Collaborative filtering depends on the user rating to construct a matrix of user-items based on the ratings of comparable customers with the same or related interest preferences \cite{larose2014discovering}. By calculating user profile similarities, the system then finds the user with similar feelings and preferences, building a comparable user group called neighborhood \cite{chulyadyo2014personalized}. CF is a well-established technique that performs well.

\begin{figure}[!ht]
 	\centering
 	\includegraphics[scale=0.4]{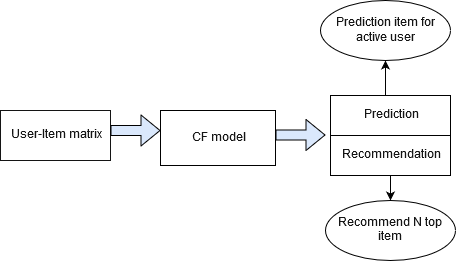}
 	\caption{Collaborative filtering process}
 	\label{Collaborative filtering process}
 \end{figure}

As such, CF enables the user to give ratings on a collection of components in such a manner that, if sufficient data is collected in the system, suggestions can be made to every user based on data supplied by those users that we think to have the most in common with them. Collaborative filtering is an excellent field of study. It depends on a rating database, then the scores are compared to the customer who gives the request \cite{polatidis2016multi}. CF-produced recommendations can be either predictive or recommended. Prediction is the value of statistics. Since Recommendation is a list of the top N items that the user will like the most as seen in Fig. \ref{Collaborative filtering process}.

 The CF method is split into two classifications: memory-based and model-based \cite{bobadilla2013recommender}.
\subsubsection{Memory-based}
The memory-based method identifies similarities between the active user and other users using similar mechanisms such as Jaccard similarity, Pearson correlation coefficient, Cosine similarity, etc. The missing rating of an active user is then estimated, and the Top-K rated product is recommended to the active user. The products that a user rated or bought in the past play a significant part in looking for a customer who shares his perception \cite{zhu2009personalized}. Once a user's neighbor is discovered, it is possible to use distinct algorithms to combine neighbor preferences to produce recommendations. They have attained significant success in real-life apps due to the efficiency of these methods. Memory-based CF is divided into two classifications: user-based and item-based. 
\begin{itemize}
   \item \textbf{User based collaborative filtering:}
    User-based filtering technique computes user similarity by using comparison with their preference to the same product and calculates the preference for active user products \cite{isinkaye2015recommendation}. Fig. \ref{User based CF} shows the user-based CF techniques.
    \begin{figure}
 	\centering
\includegraphics[width=\linewidth, height=4.9cm]{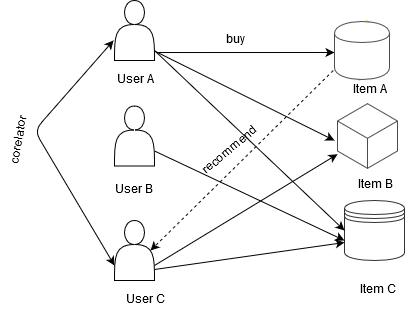}
 	\caption{User based CF}
 	\label{User based CF}
\end{figure} 
   \item \textbf{Item based collaborative filtering:}
Item-based CF calculates items similarity to make predictions. This technique first collects all items that active users buy and then detects correlations between the items and the target item. It then selects the top N items which are the most similar to determine the choice of the active user for the target item \cite{isinkaye2015recommendation}. 
\begin{figure}
    \centering
\includegraphics[width=\linewidth, height=5cm]{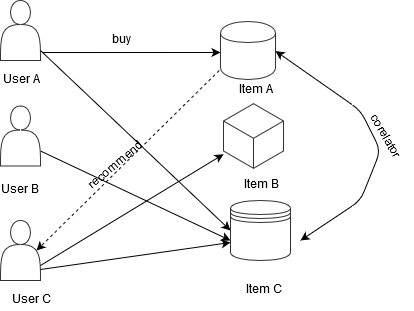}
 	\caption{Item based CF}
 	\label{Item based CF}
     \end{figure}
Fig. \ref{Item based CF} shows item-based CF techniques.
 \end{itemize}
\vspace{.2mm}
\subsubsection{Model-based}
Previous ratings are used in a model-based method to create a model using the method of machine learning. Predictions can be produced for an individual customer once the model is created. This method uses the past scores to create a model to enhance the Collaborative Filtering Technique's efficiency. Machine learning or data mining methods can be used to develop the model. These methods can rapidly recommend a set of products using pre-computed models and have shown outcomes of recommendations that are close to those of neighborhood-based recommendation methods. This is also known as a computational intelligence-based recommender system. Clustering, Bayesian network, artificial neural network, decision tree, or neural network are commonly used in recommendation systems to construct recommendation models \cite{lu2015recommender}.

\subsection{Hybrid filtering technique} 
% \label{subsec1}
A hybrid filtering system generates better efficiency by combining two or more recommendation systems. Most recommender systems use hybrid filtering that combines content-based filtering, collaborative filtering, and other methods. Hybrid methods can be applied in several ways by separating and combining collaborative filtering and content-based filtering approaches into one model. The hybrid filtering approach combines multiple recommendation strategies to simplify the system and avoid certain limitations and problems with traditional recommendation systems. Hybrid filtering systems that can manipulate and control both user preferences and content have been developed for this purpose \cite{kim2011recommender}. 

The concept behind hybrid systems is that they can offer more accurate and efficient recommendations than a single technique because another algorithm can resolve the drawbacks of one algorithm and improve the performance of recommender systems \cite{komkhao2013incremental}. However, the hybrid approach's efficiency is superior to a single approach because it reduces the method's disadvantages. To this end, these systems combine different previous system recommendation techniques to achieve some interaction between them \cite{wu2015item}. Such systems have to be closely designed so as not to inherit the drawbacks of the chosen system. 

\begin{figure}
    \centering
 	\includegraphics[width=\linewidth]{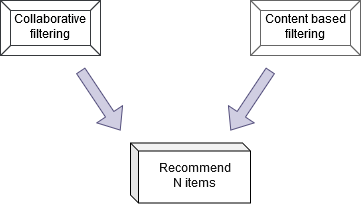}
 	\caption{CBF and CF combines}
 	\label{CBF and CF combines01}
 \end{figure}

\begin{figure}
    \centering
\includegraphics[width=\linewidth]{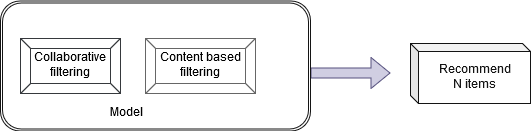}
 	\caption{CBF and CF combines}
 	\label{CBF and CF combines02}
 \end{figure}

\begin{figure}
    \centering
 	\includegraphics[width=\linewidth]{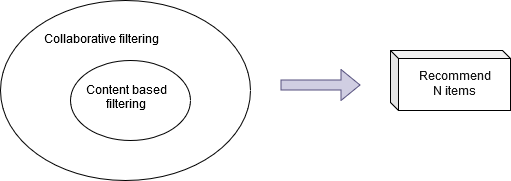}
 	\caption{CBF and CF combines}
 	\label{CBF and CF combines03}
 \end{figure}

It is possible to combine CBF and CF in various ways \cite{adomavicius2005toward}. Fig. \ref{CBF and CF combines01} demonstrates the techniques for calculating and subsequently combining CBF and CF suggestions individually. 
 In Fig. \ref{CBF and CF combines02}, the techniques that integrate CBF features into the CF strategy are shown. Add CBF projections to the CF rating matrix. Then it modifies the rating matrix entry by pairing it with the matrix produced by clustering the products by their characteristics for the CF. Authors integrate personality features into the measure of CF similarities to reduce the issue of new users \cite{hu2010using}.
Fig. \ref{CBF and CF combines03} shows the techniques used to build a unified CBF and CF model. The writers use Bayesian networks in other studies to combine the features of CBF and CF and produce more reliable suggestions. 

 \vspace{4mm}

  Another hybrid method is a combination of characteristics where characteristics obtained from distinct sources of information are combined and supplied to a single algorithm for advice. This technique enables the system to consider cooperative information without relying only on it, thus reducing the system's sensitivity to the number of customers who rated one item. Instead, it enables the system to make resemblance data about products \cite{shah2015hybrid}.
  
There are seven different types of Hybridization processes as Weighted, Cascade, Switched, Mixed, Meta-level, Feature augmentation, and Feature combination \cite{isinkaye2015recommendation} that are described on table \ref{Hybridization methods}.

% Table 01 - Start

\begin{table}[!ht]
\begin{footnotesize}
\caption{Hybridization methods}
\label{Hybridization methods}
\begin{tabular}{p{3cm}p{4cm}}
\hline\noalign{\smallskip}
    Hybridization methods&Description\\
\noalign{\smallskip} \hline
    Weighted &
The decision is taken based on the score obtained for this hybrid method from a different recommender system. The result of each recommender system is combined in a single numerical component in order to determine the final list of recommendations.
\\
Cascade &
In cascade, recommendation is based on a series of recommendations, One technique's recommendations are refined through another recommendation technique. A  list of suggestions is produced by the first recommendation method, which is improved by the next recommendation technique.
\\
Switched &
This method can be used to select One strategy in the set of recommendation strategies available
\\
Mixed &
The different recommended systems function together to provide the final personalisation list with a collaborative decision.
\\
Meta-level &
For another recommendation system, the output of a single recommendation system is used as input.
\\
Feature Combination &
Different features of the source of knowledge are combined to create a single domain.
\\
Feature Augmentation &
One information source's features are calculated to make it usable for any other recommender algorithm to function as an input
\\
\hline
\end{tabular}
\label{table:Hybridization methods}
\end{footnotesize}
\end{table}

%  Table 01 - End

\break
\subsection{ Knowledge based Recommendation Techniques} 
% \label{subsec1}
Knowledge-based recommendation systems are a specific type of recommendation system focused on explicit knowledge of the variety of items, user preferences, and criteria for recommendations. These systems are used in circumstances where content-based and collaborative filtering cannot be used. \\

The knowledge-based recommendation provides users with items based on user knowledge, items, and/or relations. In addition, a knowledge-based system provides a practical knowledge base that determines how a particular item meets the needs of a particular user, which can be done based on assumptions about a user's desires and suggestions. The knowledge base contains rich user and item information that can help generate more intuitive and better explanations for the recommended items \cite{burke2002hybrid}.

Catherine et al. \cite{catherine2017explainable}  demonstrated how the use of external knowledge in the form of knowledge graphs can produce explanations. The proposed method ranks items and knowledge graph entities in conjunction with their explanations, using a Personalized PageRank procedure to produce recommendations. The paper works on the case of film recommendations and generates a ranked list of entities as explanations by ranking them together with the corresponding films. 

\begin{figure}
    \centering
 	\includegraphics[width=\linewidth]{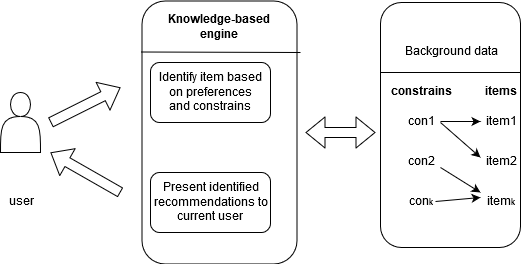}
 	\caption{Knowledge based recommendation process}
 	\label{Knowledge based recommendation process}
 \end{figure}

Unlike \cite{catherine2017explainable} who adopted guidelines and knowledge graph programming for explainable advice, Ai et al. \cite{ai2018learning} suggested embedding information graphs for explainable advice. In this study, the writers have created a user-item knowledge graph that includes a variety of user, product, and entity relationships, such as user purchased items, category items, and items jointly purchased, etc.
The knowledge-based recommendation is based on various data, such as a set of items, rules, or similarity metrics. Rules describe which items should be recommended, based on the user's demands. In terms of item property requirements, the user expresses his requirements that are both internally represented in terms of the rule. Fig. \ref{Knowledge based recommendation process} shows the KB system. \\

 \subsection{ Context-aware recommender system} 
% \label{subsec1}
Context-aware recommendation systems create more valid recommendations by applying them to the user's specific contextual condition. This recommended system works on contextual information such as weather forecasting, daytime weather forecasting, geographical location, and rain prediction. The context approach is now used by most effective recommender systems to reliably utilize information \cite{chen2014hybrid} \cite{mayer2015identifying}. 
\\\\A technique was suggested for extracting an emotional context automatically from YouTube-based films related to personal comments in the paper \cite{orellana2015mining}. Likamwa et al. \cite{likamwa2013moodscope}included speed, light, and other physical characteristics in their affective recommendation system that addresses the person's behavioral status and provides a mood for involvement in context-aware computation. A multimodal recommendation system was proposed that includes contextual data, interaction data, and users' affective behaviors and made important suggestions for unseen videos \cite{arapakis2009enriching}. \\

The context recommendation has three primary methods: pre-filtering, post-filtering, and context modeling \cite{adomavicius2011context}. These method are described on table \ref{CARS methods}  \\

% Table 02 - Start

\begin{table}[!ht]
\begin{footnotesize}
\caption{CARS methods} 
\label{CARS methods}
\begin{tabular}{p{3cm}p{4cm}}
\hline{\smallskip}
    Approaches & Description\\
\noalign{\smallskip} \hline
Context Prefiltering &
Context is used to choose any data set and then predict as a traditional system of recommendations. Another way to do this is to separate items or users by context as though they were different items or users.
\\
Context Postfiltering &
Ratings are predicted and, using the context, the results are filtered. This can be done by ordering the results according to the context or simply filtering the results
\\
Context Modeling &
The context is used in the model right. It is more complex, and multiple machine learning models such as SVM, matrix factorization, or a markov chain could implement it.
\\

\hline
\end{tabular}
\label{tab:CARS methods}
\end{footnotesize}
\end{table}

%  Table 02 - End

\section{Recommender system in e-commerce domain}

E-commerce has grown quickly in this Internet era, enabling millions of products to be sold. E-commerce customers suffer from having to choose items from millions of items. Recommender systems have become the most common method in the e-commerce industry to handle information or data overload problems. A recommendation system assists the customer in selecting products from millions of items. RS is currently being used on nearly every e-commerce website, helping millions of customers. \\

Recommender systems have helped many e-commerce platforms like Amazon.com, Netflix, Pandora, Last.fm, flikart.com, e-bay, and so on to boost their profits over the past century by using multiple Recommender systems methods. The outcome of this system is used by both companies and users of e-commerce. The recommender system not only helps the customers to get the desired product but also increases the company's income by selling more products \cite{rodrigues2016efficient}. \\

For e-commerce services, the rating is a popular feature, particularly for electronic products. In \cite{jiang2019trust}, a slope one algorithm was suggested that focused on the combination of trustworthy users and data similarities using a rating dataset. In order to fix the problem of improving the quality of recommendation systems, Qian et al. \cite{qian2019ears} develop an emotion-aware RS focused on hybrid information fusion in which three representative forms of information are merged to evaluate the characteristics of the user comprehensively user feedback data, social network data, and reviews are used respectively as explicit, implicit, and emotional information. \\

Fuzzy methods are also used in recommender systems for CB e-commerce. For products made up of distinct parts, Cao et al. \cite{cao2007intelligent} created a fuzzy based RS. When purchasing a laptop, shoppers may regard each component's personal efficiency, such as memory, motherboard, CPU, etc. The weights of shopper requirements on each element are collected and a fuzzy similarity model of the measurement model is then used to produce the most satisfied applicants. In the music recommendation system suggested to handle rich social media data, a hypergraphic model has been developed to improve the use of rich social information \cite{tan2011using}. The user recommendations should be explained in such a shopping assistant system. For instance, when buying expensive goods from intelligent sales assistants, which are able to balance the different requirements of the user, buyers expect to be skilfully pushed through the options. \\

Real-time dynamic recommendation system for all website visitors, whether they are registered or not registered \cite{lopes2015dynamic}. A rational recommendation system based on action is suggested, utilizing lexical patterns to produce recommendations for items. The efficacy of the proposed system is assessed by collecting real-time e-commerce data and comparing the system to item-based and user-based techniques. This system provides good quality accuracy and minimizes the traditional recommendation system's limitations. A context-aware learning system was developed to predict the rating known as CARL \cite{wu2019context}. For a specified user-item pair, CARL generates a joint representation based on latent interactions and features of their individual. CARL then uses Factorization Machines to model higher-order feature interactions, based on the user-item pair to predict the rating. \\
Lin et al. \cite{lin2018mulattenrec} suggest a multi-level attention system to explore the usefulness of reviews and the meaning of words and the MulAttRec Model for Recommendation. They implement a hybrid prediction layer that models the non-linear link between users and objects through the correlation Factorization Machine to the deep neural network, which highlights both high order and low order. \\

In short, for digital and physical products, e-commerce recommendation platforms are frequently used in online purchasing. To use their new algorithms, the researchers have developed several successful applications for e-commerce systems. Such systems provide developers with good guidelines on how to implement e-commerce recommender systems in practice. \\

\section{Recommender system in E-learning and E-library domain}

The use of e-learning and e-library resources has expanded significantly in recent years. Such developments have resulted in a significant increase in the availability of online learning and library services. With this increase in resources for e-learning and e-library, it's hard for learners to find useful and relevant resources. Recommended systems can solve the problem of information overload by filtering out redundant resources and choosing appropriate resources automatically for individuals based on their individual preferences. \\
A personalized suggestion model for e-learning materials (PLRS) was introduced in \cite{lu2004personalized}. Once a learning activity database is formed and the system obtains information about the registration of a learner, the PLRS uses a model for computational analysis to classify the learning requirements of an individual and then deploys matching rules to produce a recommendation. A fuzzy item response theory is introduced in a personalized recommender system for courseware designed in \cite{chen2008personalized} to initially gather the interests of a learner, after which, as a percentage of their understanding of the training courseware, the learner receives a fuzzy answer. \\
A hybrid KB recommendation system based on sequential pattern mining and ontology to advise learners on e-learning tools \cite{tarus2017hybrid}. In this proposed recommendation method, ontology is used to design and describe the learner and training resources' domain knowledge, while the sequential pattern mining algorithm discovers the sequential learning habits of the learners. This hybrid approach will overcome both the problems of data sparsity and cold starts by using the learner's sequential access approach and ontological domain knowledge. \\
A user-centric design approach to applying Web-based education systems with personalized support was proposed and applied to the Willow system, \cite{santos2014extending}. This study shows that the creation of learning personalized e-environments is a process that takes into account the needs of learners throughout the life cycle of e-learning. It also noted the use of the e-learning life cycle in developing and evaluating personalization support through feedback on Web-based education systems. A personalized online interface for learning objects offering a comparison of the learner profile and learning object explanation is proposed in the paper \cite{biletskiy2009adjustable}. The comparison is based not only on the principles of the qualities of the student profile and the explanations of the characteristics of the learning object, but also on the significance of those attributes. \\

A recommendation system in university digital libraries to recommend research resources \cite{porcel2010dealing}. In these, a Fuzzy Linguistic RS was suggested where multi-granular Fuzzy Linguistic Modeling has been used to represent and manage flexible information through language labels, and a hybrid recommendation system was presented that combines CF and CB methods. Users are allowed to designate their expectations through an ambiguous fuzzy linguistic preferential interaction to reduce the user feedback initiative. Huang et al. \cite{huang2018research} developed a self-adaptive website for an e-library recommendation system. It is split into two parts: online and offline. The former consists of data collection, preprocessing data, and periodic pattern mining. The latter creates sets of recommendations by using the university library. To get the current user access paths, a sliding window approach is used, and then the aggregation tree-based association rule algorithm is used to produce an association rule set. Finally, to get a recommendation set, you have to use the recommendation set generation algorithm. \\

The hybrid recommendation methods that combine CB, CF, and/or KB strategies are commonly used in the above-mentioned e-library recommendation systems. One reason for using hybrid methods is to take advantage of the merits of various recommendation strategies. KB plays a more significant role in making decisions for e-learning RS systems than other techniques.

\section{Recommendation system in e-tourism domain}

E-tourism recommender systems are built to provide recommendations to travelers. Many programs concentrate on attractions and locations, while others provide travel packages that include buses, accommodation, and restaurants. \\
A personalized sightseeing planning system (PSiS) was created to help tourists locate a custom tour plan in Porto, Portugal \cite{lucas2013hybrid}. The proposed hybrid recommendation approach used strategies CB and CF, fuzzy logic, and combined a clustering technique with an associative classification algorithm to improve recommendation performance. SigTur was programmed to provide personalized tourism recommendations in the Tarragona region \cite{moreno2013sigtur}. The SigTur has combined many types of information and recommendation methods to make appropriate suggestions. The recommender's information includes demographic data, specifics describing the travel background, geographic details, etc. The SigTur uses many recommendation techniques, such as CF and CB techniques, stereotypes, and artificial intelligence tools. \\
The SPETA system uses the knowledge of a client's current position, interests, and previous location history to determine the facilities that visitors receive from a professional tour guide \cite{garcia2009speta}. To enhance the interactions of travelers, this combines social networking and context-awareness techniques. It provides a customized guide to the tourist and addresses the question of collapse. A mobile-based recommendation system called SMART MUSEUM makes recommendations to users on their mobile devices for sites and particles on those sites \cite{ruotsalo2013smartmuseum}. A personalization, description, and information filtering model based on ontology has been built in this process. Contextual data is mapped to the ideas described in the ontologies, whether input by users or captured by mobile device built-in sensors. 

Herzog et al. \cite{herzog2018tourrec} created a smartphone tourist travel recommendation system named TourRec, which sequences points of interest along nice routes. TourRec's core is a modular, multi-tier architecture that facilitates the development and analysis of new recommendation algorithms, customers, and data sources. They demonstrate how the TourRec Android device can be used to guide individuals and groups on tourist trips. TourRec supports the evaluation of different recommendation algorithms and group recommendation techniques. PCAHTRS, a personal context-aware hybrid travel recommender system, is developed by integrating the contextual information of the traveler's \cite{logesh2019exploring}. PCAHTRS is evaluated on datasets of Yelp and TripAdvisor's real-time large-scale. The experimental results show that their system has improved performance over traditional approaches. This will work as a future guide to help researchers find fruitful solutions to recommendations. 

In summary, e-tourism recommendation systems employ different recommendation techniques according to the degree of complexity and their specifications. Usually applied to recommend CB, KB, and CF techniques in e-tourism. 

\section{Recommender system in e-government domain} 
% \label{sec1}
E-government is the use of technical devices, such as computers and the Internet, to provide public services in a country or region to citizens and other people. E-government offers opportunities for citizens to have more specific and efficient access to the government and to provide service to citizens. The rapid evolution of e-government has caused information overload. Recommended systems can solve this problem and have been adopted in applications of e-government. E-government recommendation services are included  G2C (Government to Citizens) and G2B (Government to Business). 
\subsection{G2C service recommendation}
% \label{subsec1}
A recommendation system known as TPLUFIB-WEB has been introduced to provide personalized exercises and preventional recommendations for low back pain patients \cite{esteban2014tplufib}. With a reduction in travel costs and staff costs, the system may be used anywhere and at any time. It is very user-friendly, designed for people with minimal skills, and uses fuzzy linguistic models to increase user preferences and encourage interactions between users and the system. De Meo et al. \cite{de2008decision}. presented a multi-agent system to support citizens' access to personalized and adapted public services. In accordance with the user profile and the device profile used, the suggested system recognizes and suggests the most interesting services for the user. The fuzzy-based RS model's preliminary results are applied to the Municipality of Quito dataset \cite{meza2019fuzzy} to recommend citizens' payment behavior. The system suggested provides some insights into using RSs to raise awareness of tax payments among people and thus improve the income of government institutions. 

\subsection{G2B service recommendation}
% \label{subsec1}
A recommendation system called BizSeeker has been developed to support the government in effectively recommending proper business clients to individual businesses \cite{lu2010bizseeker}. BizSeeker provides a list of potential business partners for business users. A method of semantic relevance for the estimation of semantic similarity was suggested, and a hybrid semantic RS system was created which combines item-based semantic similarity and CF similarity techniques. A real-world case study reveals how BizSeeker can solve the problem of sparsity and improve the accuracy of the recommendation. Business profiles usually present dynamic tree architectures and user expectations are fuzzy and ambiguous. A preference tree-based recommendation framework for personalized B2B e-services has been built and adapted to the business partner's recommendation system \cite{wu2014fuzzy}. \\

Trust or reputation is vital in business partner selection and has a major influence on the decision of a business user on whether he should do business with another partner or not. A hybrid trust-enhanced collaborative filtering RS known as the TeCF approach, which combines enhanced user-based CF and implicit trust filtering approaches, has been proposed to solve cold start and data sparsity problems to improve accuracy \cite{shambour2012trust}. \\
In addition to traditional CB and CF, fuzzy strategies that make hybrid recommendation approaches are being used to improve the performance of personalized e-government services. \\

\section{Recommendation system in e-resources domain} 
% \label{sec1}
The e-resources services relate to the content that users upload, such as images, music, files, TV programs, webpages, documents, and videos. Some users share their resources on the internet so that other users can get resources that they are interested in. In these sections, we focused on several recommendation systems applications for resource service recommendations for both individual and group users. \\
CoFoSIM, a recommendation method for mobile music, uses multi-criteria decision-making methods to evaluate partial listening records and implicit feedback and combine them into composed preferences \cite{lee2010collaborative}. The music recommendation platforms have interesting features that use some implicit feedback to improve or remove explicit ratings of users.CinemaScreen is a system of recommendations for movies \cite{salter2006cinemascreen}. One feature of the recommendation systems for film and music is that it is not easy to get the navigation history and content from multimedia sources. These resources include features such as genres, artists, etc. Combination of CF and CB methods in these film recommendations to solve cold start problems. \\

Katarya et al. \cite{katarya2018recommender} developed a new collaborative recommendation system focused on films that uses the bio-inspired gray wolf optimizer and the c-mean fuzzy clustering algorithm to calculate the movie score based on historical data and user similarity for a specific user. To get the initial clusters, the gray wolf optimizer algorithm is applied to the well-known Movielens dataset. FCM is used to classify users by similar scores in the dataset. This recommender system performed extremely well in terms of precision and accuracy. \\
Zheng et al. \cite{zheng2011recommender}  have implemented a CF-based system of folksonomy recommenders. In the recommendation process, they manipulated the tag and time effects. They developed matrixes focused on tag and time relationships instead of using the score matrix in conventional CF. \\
Three methods are used to measure the correlations dependent on matrices: tag-weight, time-weight, and mixed. In online broadcasting, Park et al. \cite{park2017rectime} introduced an RS system for real-time online broadcasting named RecTime. RecTime concurrently takes into account time factors and interests. In addition, RecTime uses a 4-d tensor factorization that takes into account two more dimensions of time variables, whereas traditional collaborative filtering approaches only take into account two dimensions, users and objects. The system automatically determines both the desired time and the products by factorizing the 4-d tensor at the same time. \\

Jalali et al. \cite{jalali2010webpum}  built an online prediction system called WebPUM and proposed a new method for classifying navigation patterns of users' browsing to predict users' future intentions. The online prediction system is efficiently implemented by using web usage mining. The navigation pattern mining method is based on the new graph partitioning algorithm to model user navigation patterns. To predict the next user phase, the most common longest subsequence algorithm is used to distinguish current user operations from previous ones. Nguyen et al. \cite{nguyen2013web} proposed that the model could navigate more effectively by combining ontology and semantic knowledge used to evaluate session results to recommend web pages to users. \\

Vildjiounaite et al. \cite{vildjiounaite2009unobtrusive} developed a TV program recommendation system for family members. This research provides a method for learning a joint model of a multi-user system from implicit interactions. Program choices are made collectively and separately by family members. The proposed method adapts to each family's behaviors and preserves the family's privacy, as the common family model is learned differently for each family. \\
In summary, the purpose of these e-resource service recommendation systems is to organize and manage Web service content of this type and to save users from boring searches. In CF e-resource services, intelligent techniques such as semantic analysis, bayesian classifier, and decision tree are combined with CB and CF approaches to implement recommendation systems.

\section{Recommender system Evaluation Metric}

Recommendation system quality is determined by how relevant they are to our interests and how unusual and interesting they are. We take a look at some of the most commonly used metrics for evaluating recommendation systems.

\subsection{Precision}

The proportion of relevant things obtained out of all the items retrieved is a measure of accuracy. Accordingly, if our recommender system selects five items to suggest to users, and three of them are related, the precision will also be 60\%. Precision includes gathering the most relevant items for the user, presuming that more beneficial products are available than you desire.

\begin{equation}\small
Precision = \frac{The\;Number\;of\;recommendation\;items\;that \\\;are\; relevant}{Number\; of\; items \;recommended}
\end{equation}

The proportion of relevant recommended items in the top-k set is called precision at k.

\begin{equation}
Prec\left ( R_{k} \right )= \frac{\left| \left\{ r\epsilon R:r\leq k\right\}\right|}{k}
\end{equation}

The number of relevant items from recommendations is r€R, while the total number of suggested items is k.

\subsection{Recall}

The proportion of relevant things retrieved from all relevant items is a measure of completeness. If there are five relevant things and the recommender chooses two, the recall will be 40\%. The purpose of the recall is to ensure that important things do not go missing.

\begin{equation}\small
Recall = \frac{The\;Number\;of\;recommendation\;items\;that\;are\; relevant}{Number\; of\; all \;possible\;relevant\;items}
\end{equation}

The proportion of relevant things discovered in the top-k recommendations is known as recall at k.

\begin{equation}
Recall\left ( R_{k} \right )= \frac{\left| \left\{ r\epsilon R:r\leq k\right\}\right|}{R}
\end{equation}

The number of relevant things from recommendations is r€R, and the number of relevant items for the provided item is $\left|R \right|$.

\subsection{Average Precision (AP)}

Average Precision (AP) is a ranking precision metric that prioritizes correct prediction with a high rank. Precision can help to evaluate the model's overall performance, but it can't tell you if the products were appropriately rated. The AP aids in determining the quality of the recommender model's rating of the selected item. 
If we really need to recommend N items and there are m relevant items in the entire object space, we can say:

\begin{equation}
  \begin{aligned}[b]
      AP@N=\frac{1}{m}\;\sum_{k=1}^{N} (P(k)\; if \; k^{th}\; item \;was\; relevant )\\ =\frac{1}{m} \sum_{k=1}^{N} \;P(k) \cdot rel(k)
  \end{aligned}
\end{equation}

Where P(k) is the precision value at the $k^{th}$ recommendation and rel (k) is whether the $k^{th}$ recommendation was relevant (rel (k) = 1) or not (rel (k) = 0).

\subsection{Mean of Average Precision (MAP)}

The mean average precision is MAP. When we have the AP for each user, calculating the MAP is as simple as averaging it over all users. We've only computed the AP for one user; to determine the system's overall performance, we'll need to compute the AP for each user. We have |U| APs, so we take the average once more to get a single measure. Because we are taking the average of precisions computed up to K and computing the mean of average precision@K for each user to obtain the overall system performance.

\begin{equation} \small
MAP@N \\=\frac{1}{\left| U \right|}\;\sum_{u=1}^{|} U \left| (AP@N)_{u }= \frac{1}{\left| U \right|  } \sum_{u=1}^{|} U \right | \frac{1}{m} \sum_{k=1}^{N} P_{u} (k) \cdot rel_{u} (k) 
\end{equation}

We are adding up the AP@N for each user normalized by the number of users, and |U| is the number of users.

\subsection{F1-Measure (FM)}
This measure represents the harmonic mean of recall and precision. The F1 metric considers both precision and recall and creates a compromise between the two. The F1 measure's formula is as follows:

\begin{equation}
F1-Measure = 2\ast \frac{Preccision\ast Reccall}{Preccision + Reccall}
\end{equation}

\subsection{Receiver Operating Characteristics(ROC) Curve}

The ROC curve can help you determine what threshold will give you the best results. That's how it looks graphically. The ROC curve represents the relationship between properly predicted items (TPR) and wrongly predicted items (FPR). It gives information if the purpose is to fine-tune the recommender's effectiveness by identifying its targets.

\begin{itemize}
    \item  True Positives (TP): Actually Positive and Positive Prediction.
\item True Negative (TN): Actually Negative and Negative Prediction.
\item False Positive (FP): Actually Negative and Positive Prediction.
\item False Negative (FN): Actually Positive and Negative Prediction.
\end{itemize}

On this curve, two parameters are plotted:
\begin{itemize}
    \item  The term "True Positive Rate (TPR)” is a synonym for "recall," and it is defined as follows:
    \begin{equation}
TPR =  \frac{TP}{TP + FN}
\end{equation}

 \item The following is how the False Positive Rate (FPR) is defined:
 \begin{equation}
FPR =  \frac{FP}{FP + TN}
\end{equation}
\end{itemize}

\begin{figure}
    \centering
 	\includegraphics[scale=0.5]{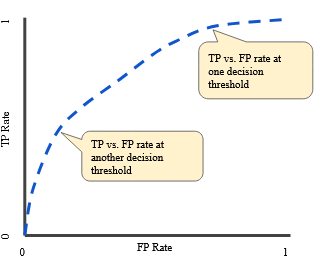}
 	\caption{ROC Curve}
 	\label{ROCcurve}
 \end{figure}

The graph (see Fig. \ref{ROCcurve}) depicts a typical ROC curve.

TPR vs. FPR at various classification thresholds are plotted on a ROC curve. When the classification threshold is lowered, more items are classified as positive, resulting in an increase in both False Positives and True Positives.
\subsection{Area Under ROC Curve (AUC)}
The probability of a random relevant item being ranked higher than a random irrelevant one is measured by the AUC. The higher the frequency of this occurring, the higher the AUC value, and hence the better the recommendation system. In other words, AUC evaluates the entire two-dimensional area beneath the entire ROC curve from (0,0) to (1,1). 
\begin{figure}
    \centering
 	\includegraphics[scale=0.6]{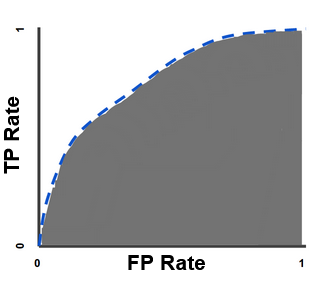}
 	\caption{AUC Curve}
 	\label{AUCcurve}
 \end{figure} 
A typical AUC curve is depicted in Fig. \ref{AUCcurve}.

\subsection{Mean Absolute Error (MAE)}

The mean absolute error is the average of the difference between the value predicted by the recommender and the actual value given by the user. The MAE shows how close the predicted score is to the actual score. The MAE of 0 implies that the expected and actual ratings were identical, suggesting that the model successfully predicted. As a result, a lower MAE is preferable.
\begin{equation}
MAE=\frac{\sum_{i=1}^{n}\left|y_{i}-x_{i} \right|}{n}
\end{equation}
As a result, it's an arithmetic average of absolute errors, where $y_{i}$ the prediction and $x_{i}$ is the actual value.

\subsection{Mean Squared Error (MSE)}

The fundamental distinction between Mean Squared Error and Mean Absolute Error is that instead of negating the negative sign with the absolute error, we square it. Because MAE penalizes results, even a small difference can make a big difference. This also implies that the recommender system performed admirably if the MSE is close to zero.

\begin{equation}
MSE=\frac{1}{n}\sum_{i=1}^{n}\left ( Y_{i}-\hat{Y}_{i} \right )^{2}
\end{equation}
Here, $Y_{i}$ is the vector of observed values for the variable to be predicted, and  $\hat{Y}_{i}$   is the predicted values.

\subsection{Root Mean Square Error (RMSE)}
The Root Mean Square Error (RMSE) is a standard method for estimating a model's error in predicting quantitative data. RMSE is similar to MAE but places more emphasis on larger deviations. Although MSE helps to negate the negative sign, it still scales up errors that really are pointless to compare to actual rating values due to differing rating scales. To normalize the scale issue that MSE had, we take the square root of MSE in RMSE. The RMSE will nearly always outnumber the MAE. To act as a heuristic for training models and to assess the utility and correctness of trained models.
It is formalized as follows:
\begin{equation}
RMSE=\sqrt{\sum_{i=1}^{n}\frac{\left (\hat {y}_{i}-{y}_{i} \right )^{2}}{n}} 
\end{equation}
Here, $y_{i}$ is  observed values for the variable to be predicted, and  $\hat{y}_{i}$   is the predicted values. And n is the number of observations.

\subsection{User Coverage}

By establishing a threshold that allows only excellent recommendations for each user in our top n list, adding them, and dividing them by the number of users in our top n list, we hope to find good recommendations in our top n list. Our recommender's coverage is 70\% if we have 10,000 things and the model covers 7,000 of them for various people. If the goal is to provide the user with as many options as possible, coverage may be a valuable tool for assessing the recommender model.
\begin{equation}
Coverage=\frac{\left| \bigcup _{u\epsilon U}R_{u} \right|}{\left| I \right|}
\end{equation}

Where U is the set of all system users, and I is the item catalog. $R_{u}$ is the set of all recommendations given for user u.

\subsection{Novelty}

Novelty contributes to the understanding of the model's behavior. Users will be more interested in the same things that they purchased. Therefore, the novelty may not be effective when proposing items to them at checkout. However, a place where the customer is still browsing the website and proposes something wholly new and unique might be valuable, and novelty can assist in quantifying that. Comparing a user's suggested products to the population is a typical way of evaluating novelty. This may be measured in two ways.

\begin{equation}
Novelty\left ( i \right )= -log_{2}\;\frac{count\;  \left ( users\;recommended\; \textit{i} \right )}{count\;\left ( all \;users \right )}
\end{equation}

% Equ 15
\begin{equation}
Novelty\left ( i \right )= 1-\;\frac{count\;  \left ( users\;recommended\; \textit{i} \right )}{count\;\left ( all \;users \right )}
\end{equation}
Here $ \textit{i}$ is the item.

\subsection{Diversity}
Similar to novelty, understanding our model's diversity is helpful depending on the domain and where we are to offer products. It's helpful to know how diverse our model's recommendations are. As a result of our high diversity, our users will always have something new and interesting to see and consume. This is determined by first determining user similarity and then subtracting it from 1 to determine diversity.

\begin{equation}
D=\frac{\sum_{i=1}^{n}\sum_{j=1}^{n}\left ( 1-Similarity\left ( c_{i},c_{j} \right ) \right )}{n/2\ast \left ( n-1 \right )}
\end{equation}

\subsection{Unexpectedness}

Unexpectedness recommends items to users that are different from what they may expect from the system. It may be used to keep track of marginal progress in recommendation systems. One method is to compare recommendations to the user's previous item interaction history. Another option is to utilize a distance metric to calculate the cosine similarity between a user's recommended items (I) and previous item interactions (H). Higher unexpectedness is indicated by a lower cosine similarity.

\begin{equation}
Unexpectedness\left ( I,H \right )=\frac{1}{I}\sum_{i\epsilon I}\sum_{h\epsilon H} Cosine Similarity\left ( i,h \right )
\end{equation}

\subsection{Serendipity}

Serendipity is a criteria for producing recommendations that are both attractive and useful. The key advantage of this criteria over novelty and diversity is the utility of serendipitous recommendations. Unexpectedness multiplied by relevance, where relevance is 1 if the recommended item is interacted with and 0 otherwise, is how serendipity is calculated. We only evaluate the unexpectedness of a recommended item (i) if the user engages with i.
\begin{equation}
Serendipity(i)= Unexpectedness(i)\times  Relevance(i)
\end{equation}
Where Relevance(i)=1 if i is interacted upon, else 0. To get overall Serendipity, we average over all users (U) and all recommended items (I).

\begin{equation}
Serendipity=\frac{1}{count\left ( U \right )}\sum_{u\epsilon U}\sum_{i\epsilon I} \frac{Serendipity(i)}{count(I)}
\end{equation}

\subsection{Normalized Discounted Cumulative Gain (NDCG)}

The NDCG has three parts. Initial Cumulative Gains (CG).A relevance score is assigned to each recommendation. The total of all the relevance scores in a recommendation set is the cumulative gain.
\begin{equation}
CG(k)=\sum_{i=1}^{k}G_{i}
\end{equation}

Here $G_{i}$ is the gain.\\
Discounted cumulative gain (DCG) penalizes highly relevant items that appear lower in the search results by lowering the graded relevance value by a logarithmic proportional to the result position.

\begin{equation}
DCG(k)=\sum_{i=1}^{k}\frac{G_{i}}{log_{2}(i+1)}
\end{equation}
Because the denominator is log(i+1), the items indicated at the top are given greater weight.

The DCG has one weakness: the score is determined by the number of items recommended. It's difficult to compare the DCG score of two recommenders who each recommend a different quantity of items. IDCG is a solution to this problem (ideal DCG). IDCG is the DCG score for the most ideal ranking, which is a top-to-bottom ranking of items based on their relevance up to position k.
\begin{equation}
IDCG(k)=\sum_{i=1}^{\left| I(k)\right|}\frac{G_{i}}{log_{2}(i+1)}
\end{equation}
Where I(k) representsthe ideal list of items up to position k, $\left| I(k)\right|=k$.

The purpose of NDCG is to normalize the DCG score by IDCG such that its value is always between 0 and 1, regardless of duration. Normal Discount Cumulative Gain is defined as the total of gains up to position k in the recommendation list.

\begin{equation}
NDCG(k)=\frac{DCG(k)}{IDCG(k)}
\end{equation}

\section{Recommender system challenges and issues}

There are some major research challenges and issues related to the recommender system that are discussed below:

\begin{itemize}
\item \textbf{Cold start:}
A cold start refers to the case where recommendations for a new customer and products are difficult to create. Recommendation to a new customer is difficult because he/she has not yet rated any products, and as such, there is much less data about customers. Similarly, recommending a new item is also difficult. Due to a lack of adequate rating information, it is difficult to find similarities between customers and products. Therefore, the taste of the new users cannot be predicted and the new items will not be rated or purchased. This condition decreases the recommendation system's accuracy.

\item \textbf{Scalability:}
As there are millions of internet users and products, recommendation systems face major difficulties in managing enormous amounts of data. The calculation of recommendations thus increases exponentially, becomes costly, and sometimes leads to incorrect outcomes. If the dataset size increases with the number of users and objects, the calculation also increases linearly. The algorithm works well in a small dataset but cannot produce a satisfactory outcome for a big data set quantity. Thus, applying the recommendation method to enormous and dynamic data sets generated by communication between item users is very hard.

\item \textbf{Sparsity: }
Most users have not been involved in rating most products, resulting in a generally sparse rating matrix and making it difficult to locate related users. This is an issue that happens when most consumers do not rate most products, and therefore the matrix of customer products becomes very sparse. So, there is a decrease in the possibility of having several customers of the same quality.

\item \textbf{Synonymy: }
Similar objects with distinct names or entries are referred to as synonymy. RS algorithms can not distinguish between closely associated products like "action films" and "action movies". Extreme use of synonymous phrases reduces recommendation system efficiency. Using these techniques can solve the synonymous issues of Singular Value Decomposition (SVD) and Latent Semantic Indexing.

\item \textbf{Flexibility:}
Flexibility is the capacity of a recommended system to interact in a timely and cost-effective way with potential inner or external changes. Recommendation systems techniques should be flexible for ease of modification and improvement \cite{navimipour2014resource}.

\item \textbf{Shilling Attack:}
A recommendation system is a commercial activity so that people become biased in their feedback and offer millions of positive reviews for their products but negative reviews for their competitors' products and items. The system, therefore, needs to incorporate some sort of mechanism to discourage this kind of trend.

\item \textbf{Gray Sheep:}
The user whose thoughts are not consistent with any group of people is referred to as the gray sheep. Such applications do not help collaborative filtering run smoothly. On the other hand, there is a particular class of users known as Black Sheep whose unconventional conduct makes recommendations almost impossible. A hybrid system combining CB and CF gray sheep problems can be solved.
\item \textbf{Privacy issue:}
A recommendation algorithm needs user demographic input to generate personalized suggestions of quality; this can lead to data privacy and safety problems. Therefore, a technique can be designed so that user data is used decently and carefully to insure that user information is not accessible to malicious users.

\end{itemize}
\section{Comprehensive analysis}

Recommender systems are discussed above and summarized in this chapter. The following significant conclusions can be drawn from the overview of recommended systems: \\

\begin{enumerate}
\item Traditional recommendation methods, such as CB, CF, and KB, continue to play a dominant position in almost all types of applications, but hybrid recommendation systems are more common than single recommendation-based technique schemes to avoid the disadvantages of individual recommendation methods.
\item Some model-based CF methods, such as neural network and fuzzy logic, have been implemented to address multiple uncertainties in all types of recommender systems implementation fields.

\item  Several new recommendation schemes implementation platforms, such as TV and radio platforms, have only lately appeared.
 \item The recommendation techniques implemented in a system of recommendations are limited by the system's data sources. Numerous sizes of information can be acquired in the big data age, which is useful for more accurate and comprehensive modeling of user preferences. It is predicted that more recommended system apps will be created by using large amounts of information effectively and efficiently.
 
 \item  Although many study attempts have been made to address data sparsity issues in recommender systems, in many implementation fields, this issue is still not fixed well.
 \item We found that the number of papers released on the collaborative and content-based processing techniques in recommendation systems was very large. Moreover, most of the papers have been published in renowned newspapers. In journals and conferences, respectively, Springer and Elsevier have the most published papers.
\end{enumerate}

While research in the development of the recommendation system is increasing day by day, a major issue is how to apply recommendation methods in real-world systems and how to solve the problem of large and dynamic data sets. Sometimes an algorithm works well when tested offline on a small dataset, but it becomes inefficient when used on large real-world datasets. Moreover, recommendation systems generate personal recommendations by extracting user data. This means that this user data must be protected from unauthorized access.The Summary of the systems developed in recommender system, the techniques used are summarized in table \ref{Summary of the systems developed in recommender system, the techniques used} \\

\section{Conclusion}

Several algorithms and techniques of optimization have already been proposed to improve the performance of the RS system over the past few decades. But the recommendation techniques are associated with several issues that still require the researcher's attention. Future work suggests the use of hybrid techniques with bio-inspired approaches, improving existing methods and algorithms, machine learning, and profound learning to deliver promising results. 

%\twocolumn
\begin{footnotesize}
\onecolumn
\begin{longtable}[!ht]{p{2.2cm}p{2.5cm}p{2.5cm}p{4cm}p{2.2cm}p{1.5cm}}
\caption{Summary of the systems developed in recommender system, the techniques used\label{long}}
\label{Summary of the systems developed in recommender system, the techniques used}
\\ \hline{\smallskip}
System & Techniques & Data & Evaluation & Application Platform & Year\\
\noalign{\smallskip} \hline
 \endfirsthead

\caption{Continued From previous page}\\
\hline\noalign{\smallskip}
 System&Techniques&Data&Evaluation&Application Platform&Year\\
\noalign{\smallskip} \hline
 \endhead

 \hline
 \endfoot

 \hline

 \endlastfoot
    OMLCF \cite{anitha2021optimized} & CF, SVM & Taobao dataset & Recall, F-measure, Predictive accuracy & E-commerce & 2021\\ 
    
    Fuzzy Recommender System \cite{karthik2021fuzzy} & Fuzzy logic, sentimentanalysis & Amazon review dataset & precision, recall, Diversity, Novelty and Serendipity, nDCG & E-commerce & 2021\\ 
    
    EARS  \cite{qian2019ears} & Hybrid information fusion, matrix, factorization & Rating, user social network, user review & MAP, F-measure & E-commerce & 2019\\ \\

    CARL \cite{wu2019context}  & Context aware, neural network & Reviews data, user item rating & MSE  & E-commerce & 2019\\ 
    
    MulAtten Rec \cite{lin2018mulattenrec} & Hybrid, factorization machine, Deep Neural Network & User review, rating & RMSE & E-commerce & 2018\\ 
    
    DRS \cite{lopes2015dynamic}  & Web usage mining, Product based, action based techniques & Click stream data & Recall, Precision & E-commerce & 2015 \\ 
    
    MRH \cite{tan2011using}  & Unified hypergraph & Music data & AP, NDCC & E-commerce & 2011 \\
    
    Consumer electronic products RS  \cite{cao2007intelligent} & KB, fuzzy techniques & Laptop data & Recall, Precision, F-measure & E-commerce & 2007 \\ 
    
    LRRSS \cite{huang2018research}  & Adaptive algorithm, Data mining & User browsing history data & Recall, precision, coverage & E-library & 2018\\
    
    University digital library RS \cite{porcel2010dealing} & Hybrid, fuzzy inguistic modeling &  User interactions data & Recall, Precision, F-measure & E-library & 2010\\
    
    HKBR   \cite{tarus2017hybrid} & KB, CF, spm algorithm & Ratings, web logs data & Recall, Precision & E-learning & 2017\\
    
    MoodleREC \cite{de2020moodlerec} & Hybrid Filtering & Moodle Log database & Recall, Precision, F1-Measure & E-learning  & 2020 \\
    
    Willow system \cite{santos2014extending} & KB & User interactions data & SD, SUS score & E-learning & 2014 \\
    
    PSDLO  \cite{biletskiy2009adjustable} & KB, CB & Web data, personal interests & Weights comparison & E-learning & 2009\\
    
    E-Tourism Package Recommendation \cite{kolahkaj2020hybrid} & Hybrid context, squential patter mining & Flickr dataset & Precision, MAP, AP, Recall, F1-measure & E-tourism  & 2020\\
    
    PCA- HTRS \cite{logesh2019exploring} & Context aware techniques & Yelp and TripAdvisor data &  RMSE, coverage, F-measure & E-tourism & 2019\\
    
    TourRec  \cite{herzog2018tourrec}  & Group techniques, Dĳkstra’s algorithm & Ratings, user’s feedback data & A/B testing & E-tourism & 2018\\
    
    SigTur  \cite{moreno2013sigtur} & CF, CB, context aware & Demo- graphic information, ratings & Spearman correlation & E-tourism  & 2013 \\
    
    SMART MUSEUM \cite{ruotsalo2013smartmuseum} & Ontology, context information filtering & Context, sensor data & MAP, Recall, Precision & E-tourism & 2013\\
    
    SPETA   \cite{garcia2009speta} & Semantic web, context aware & Locations data, user preferences & Recall, Precision & E-tourism & 2009\\
    
    E-govern- ment Deep RS \cite{wang2020government} & Deep neural network, random forest algorithm & 
    My Ningxia dataset & Recall, F1-Measure & E-govern- ment & 2020 \\
    
    FBDRS  \cite{meza2019fuzzy} & Fuzzy logic, knowledge based & Tax payment, Municipal tax data & Model Simulation & E-government & 2019\\
    
    TBRS  \cite{wu2014fuzzy} & fuzzy set techniques, tree matching & Australian business data , MovieLens data & Recall, Precision, F-measure & E-government & 2014 \\
    
    TPLUFIBWEB\cite{esteban2014tplufib}  & CB, CF, Fuzzy linguistic modeling, hybrid & Multi- media data & Weight over & E-government & 2014\\
    
    TSF \cite{shambour2012trust} & CF, Semantic filtering, Information fusion & Movie Lens data, Yahoo! Webscope R4 data & MAE, Coverage & E-government & 2012 \\
    
    BizSeeker \cite{lu2010bizseeker} & CF, hybrid & Products taxonomy, description, user rating & Weight factor, Prediction value & E-government & 2010\\ 
    
    DTRS   \cite{zhang2021dynamic} & Tensor factorization, polynomial spline approximation & IRI marketing data, Last.fm data & RMSE, MAE, PICP & E-resources & 2021\\
    
    MovieWatch  \cite{zhang2019personalized} & Weighted Slope One-VU, CF, k-means & MovieLens Dataset & RMSE  & E-resources & 2019\\
    
    RecTime   \cite{park2017rectime}  & CF, 4-d tensor factorization & Watch log data, Electronic Program Guide data & Timing precision, Recall, F $\alpha$ score & E-resources & 2017 \\
    
    NMCR  \cite{katarya2018recommender}  & CF, gray wolf optimizer, FCM & Movielens dataset & MAE, Recall, Precision & E-resources & 2016\\
    
    WebPage Recommender \cite{nguyen2013web} & KB, domainontology & Web log data & Satisfaction, Precision & E-resources & 2013\\
    
    Tag recommender system \cite{zheng2011recommender} & CF & Ratings, Tag, Time & Traditional log based model & E-resources & 2011\\
    
    WebPUM \cite{jalali2010webpum}  & Graph based clustering & Web log data & MinFreq, coverage, F1-measure & E-resources & 2010 \\
     
    CoFoSIM   \cite{lee2010collaborative}  & CF, ebusage mining & Clickstream, log data & Recall, Precision, F-measure & E-resources & 2010\\ 
     
    TV program recommender \cite{vildjiounaite2009unobtrusive} & Machine learning, Classifier & Viewing data & Recall  & E-resources & 2009

 \end{longtable}
 \end{footnotesize}
\twocolumn
%\section*{Acknowledgement}

%\section*{Appendixes}

\bibliography{source.bib}

\begin{thebibliography}{10}

\bibitem{adomavicius2005toward}
Gediminas Adomavicius and Alexander Tuzhilin.
\newblock Toward the next generation of recommender systems: A survey of the
  state-of-the-art and possible extensions.
\newblock {\em IEEE transactions on knowledge and data engineering},
  17(6):734--749, 2005.

\bibitem{adomavicius2011context}
Gediminas Adomavicius and Alexander Tuzhilin.
\newblock Context-aware recommender systems.
\newblock In {\em Recommender systems handbook}, pages 217--253. Springer,
  2011.

\bibitem{ai2018learning}
Qingyao Ai, Vahid Azizi, Xu~Chen, and Yongfeng Zhang.
\newblock Learning heterogeneous knowledge base embeddings for explainable
  recommendation.
\newblock {\em Algorithms}, 11(9):137, 2018.

\bibitem{anitha2021optimized}
J~Anitha and M~Kalaiarasu.
\newblock Optimized machine learning based collaborative filtering (omlcf)
  recommendation system in e-commerce.
\newblock {\em Journal of Ambient Intelligence and Humanized Computing},
  12(6):6387--6398, 2021.

\bibitem{arapakis2009enriching}
Ioannis Arapakis, Yashar Moshfeghi, Hideo Joho, Reede Ren, David Hannah, and
  Joemon~M Jose.
\newblock Enriching user profiling with affective features for the improvement
  of a multimodal recommender system.
\newblock In {\em Proceedings of the ACM International Conference on Image and
  Video Retrieval}, page~29. ACM, 2009.

\bibitem{bauer2014recommender}
Josef Bauer and Alexandros Nanopoulos.
\newblock Recommender systems based on quantitative implicit customer feedback.
\newblock {\em Decision Support Systems}, 68:77--88, 2014.

\bibitem{biletskiy2009adjustable}
Yevgen Biletskiy, Hamidreza Baghi, Igor Keleberda, and Michael Fleming.
\newblock An adjustable personalization of search and delivery of learning
  objects to learners.
\newblock {\em Expert Systems with Applications}, 36(5):9113--9120, 2009.

\bibitem{bobadilla2013recommender}
Jes{\'u}s Bobadilla, Fernando Ortega, Antonio Hernando, and Abraham
  Guti{\'e}rrez.
\newblock Recommender systems survey.
\newblock {\em Knowledge-based systems}, 46:109--132, 2013.

\bibitem{burke2002hybrid}
Robin Burke.
\newblock Hybrid recommender systems: Survey and experiments.
\newblock {\em User modeling and user-adapted interaction}, 12(4):331--370,
  2002.

\bibitem{cao2007intelligent}
Yukun Cao and Yunfeng Li.
\newblock An intelligent fuzzy-based recommendation system for consumer
  electronic products.
\newblock {\em Expert Systems with Applications}, 33(1):230--240, 2007.

\bibitem{catherine2017explainable}
Rose Catherine, Kathryn Mazaitis, Maxine Eskenazi, and William Cohen.
\newblock Explainable entity-based recommendations with knowledge graphs.
\newblock {\em arXiv preprint arXiv:1707.05254}, 2017.

\bibitem{champiri2015systematic}
Zohreh~Dehghani Champiri, Seyed~Reza Shahamiri, and Siti Salwah~Binti Salim.
\newblock A systematic review of scholar context-aware recommender systems.
\newblock {\em Expert Systems with Applications}, 42(3):1743--1758, 2015.

\bibitem{chen2008personalized}
Chih-Ming Chen and Ling-Jiun Duh.
\newblock Personalized web-based tutoring system based on fuzzy item response
  theory.
\newblock {\em Expert systems with applications}, 34(4):2298--2315, 2008.

\bibitem{chen2014hybrid}
Wei Chen, Zhendong Niu, Xiangyu Zhao, and Yi~Li.
\newblock A hybrid recommendation algorithm adapted in e-learning environments.
\newblock {\em World Wide Web}, 17(2):271--284, 2014.

\bibitem{chulyadyo2014personalized}
Rajani Chulyadyo and Philippe Leray.
\newblock A personalized recommender system from probabilistic relational model
  and users’ preferences.
\newblock {\em Procedia Computer Science}, 35:1063--1072, 2014.

\bibitem{de2020moodlerec}
Carlo De~Medio, Carla Limongelli, Filippo Sciarrone, and Marco Temperini.
\newblock Moodlerec: A recommendation system for creating courses using the
  moodle e-learning platform.
\newblock {\em Computers in Human Behavior}, 104:106168, 2020.

\bibitem{de2008decision}
Pasquale De~Meo, Giovanni Quattrone, and Domenico Ursino.
\newblock A decision support system for designing new services tailored to
  citizen profiles in a complex and distributed e-government scenario.
\newblock {\em Data \& Knowledge Engineering}, 67(1):161--184, 2008.

\bibitem{esteban2014tplufib}
Bernab{\'e} Esteban, {\'A}lvaro Tejeda-Lorente, Carlos Porcel, Manolo Arroyo,
  and Enrique Herrera-Viedma.
\newblock Tplufib-web: A fuzzy linguistic web system to help in the treatment
  of low back pain problems.
\newblock {\em Knowledge-Based Systems}, 67:429--438, 2014.

\bibitem{florea2017spark}
Adrian-C{\u{a}}t{\u{a}}lin Florea, John Anvik, and R{\u{a}}zvan Andonie.
\newblock Spark-based cluster implementation of a bug report assignment
  recommender system.
\newblock In {\em International Conference on Artificial Intelligence and Soft
  Computing}, pages 31--42. Springer, 2017.

\bibitem{garcia2009speta}
Angel Garc{\'\i}a-Crespo, Javier Chamizo, Ismael Rivera, Myriam Mencke, Ricardo
  Colomo-Palacios, and Juan~Miguel G{\'o}mez-Berb{\'\i}s.
\newblock Speta: Social pervasive e-tourism advisor.
\newblock {\em Telematics and informatics}, 26(3):306--315, 2009.

\bibitem{herzog2018tourrec}
Daniel Herzog, Christopher La{\ss}, and Wolfgang W{\"o}rndl.
\newblock Tourrec: a tourist trip recommender system for individuals and
  groups.
\newblock In {\em Proceedings of the 12th ACM Conference on Recommender
  Systems}, pages 496--497. ACM, 2018.

\bibitem{hu2010using}
Rong Hu and Pearl Pu.
\newblock Using personality information in collaborative filtering for new
  users.
\newblock {\em Recommender Systems and the Social Web}, 17, 2010.

\bibitem{huang2018research}
Zhenzhen Huang, Tianxu Li, and Shuo Xiao.
\newblock Research on library recommendation reading service system based on
  adaptive algorithm.
\newblock {\em Wireless Personal Communications}, 102(2):1963--1977, 2018.

\bibitem{isinkaye2015recommendation}
FO~Isinkaye, YO~Folajimi, and BA~Ojokoh.
\newblock Recommendation systems: Principles, methods and evaluation.
\newblock {\em Egyptian Informatics Journal}, 16(3):261--273, 2015.

\bibitem{jalali2010webpum}
Mehrdad Jalali, Norwati Mustapha, Md~Nasir Sulaiman, and Ali Mamat.
\newblock Webpum: A web-based recommendation system to predict user future
  movements.
\newblock {\em Expert Systems with Applications}, 37(9):6201--6212, 2010.

\bibitem{jiang2019trust}
Liaoliang Jiang, Yuting Cheng, Li~Yang, Jing Li, Hongyang Yan, and Xiaoqin
  Wang.
\newblock A trust-based collaborative filtering algorithm for e-commerce
  recommendation system.
\newblock {\em Journal of Ambient Intelligence and Humanized Computing},
  10(8):3023--3034, 2019.

\bibitem{jiang2015author}
Shuhui Jiang, Xueming Qian, Jialie Shen, Yun Fu, and Tao Mei.
\newblock Author topic model-based collaborative filtering for personalized poi
  recommendations.
\newblock {\em IEEE transactions on multimedia}, 17(6):907--918, 2015.

\bibitem{karthik2021fuzzy}
RV~Karthik and Sannasi Ganapathy.
\newblock A fuzzy recommendation system for predicting the customers interests
  using sentiment analysis and ontology in e-commerce.
\newblock {\em Applied Soft Computing}, 108:107396, 2021.

\bibitem{katarya2018recommender}
Rahul Katarya and Om~Prakash Verma.
\newblock Recommender system with grey wolf optimizer and fcm.
\newblock {\em Neural Computing and Applications}, 30(5):1679--1687, 2018.

\bibitem{kim2011recommender}
Hyun-Tae Kim, Jong-Hyun Lee, and Chang~Wook Ahn.
\newblock A recommender system based on interactive evolutionary computation
  with data grouping.
\newblock {\em Procedia Computer Science}, 3:611--616, 2011.

\bibitem{kolahkaj2020hybrid}
Maral Kolahkaj, Ali Harounabadi, Alireza Nikravanshalmani, and Rahim
  Chinipardaz.
\newblock A hybrid context-aware approach for e-tourism package recommendation
  based on asymmetric similarity measurement and sequential pattern mining.
\newblock {\em Electronic Commerce Research and Applications}, 42:100978, 2020.

\bibitem{komkhao2013incremental}
Maytiyanin Komkhao, Jie Lu, Zhong Li, and Wolfgang~A Halang.
\newblock Incremental collaborative filtering based on mahalanobis distance and
  fuzzy membership for recommender systems.
\newblock {\em International Journal of General Systems}, 42(1):41--66, 2013.

\bibitem{kotsogiannis2017directed}
Ios Kotsogiannis, Elena Zheleva, and Ashwin Machanavajjhala.
\newblock Directed edge recommender system.
\newblock In {\em Proceedings of the Tenth ACM International Conference on Web
  Search and Data Mining}, pages 525--533. ACM, 2017.

\bibitem{kuzelewska2011advantages}
Urszula Ku{\.z}elewska.
\newblock Advantages of information granulation in clustering algorithms.
\newblock In {\em International Conference on Agents and Artificial
  Intelligence}, pages 131--145. Springer, 2011.

\bibitem{larose2014discovering}
Daniel~T Larose and Chantal~D Larose.
\newblock {\em Discovering knowledge in data: an introduction to data mining}.
\newblock John Wiley \& Sons, 2014.

\bibitem{lee2010collaborative}
Seok~Kee Lee, Yoon~Ho Cho, and Soung~Hie Kim.
\newblock Collaborative filtering with ordinal scale-based implicit ratings for
  mobile music recommendations.
\newblock {\em Information Sciences}, 180(11):2142--2155, 2010.

\bibitem{li2015accurate}
Xiu Li, Huimin Wang, and Xinwei Yan.
\newblock Accurate recommendation based on opinion mining.
\newblock In {\em Genetic and Evolutionary Computing}, pages 399--408.
  Springer, 2015.

\bibitem{likamwa2013moodscope}
Robert LiKamWa, Yunxin Liu, Nicholas~D Lane, and Lin Zhong.
\newblock Moodscope: Building a mood sensor from smartphone usage patterns.
\newblock In {\em Proceeding of the 11th annual international conference on
  Mobile systems, applications, and services}, pages 389--402. ACM, 2013.

\bibitem{lin2018mulattenrec}
Zhipeng Lin, Wenjing Yang, Yongjun Zhang, Haotian Wang, and Yuhua Tang.
\newblock Mulattenrec: A multi-level attention-based model for recommendation.
\newblock In {\em International Conference on Neural Information Processing},
  pages 240--252. Springer, 2018.

\bibitem{logesh2019exploring}
R~Logesh and V~Subramaniyaswamy.
\newblock Exploring hybrid recommender systems for personalized travel
  applications.
\newblock In {\em Cognitive informatics and soft computing}, pages 535--544.
  Springer, 2019.

\bibitem{lopes2015dynamic}
Prajyoti Lopes and Bidisha Roy.
\newblock Dynamic recommendation system using web usage mining for e-commerce
  users.
\newblock {\em Procedia Computer Science}, 45:60--69, 2015.

\bibitem{lu2004personalized}
Jie Lu.
\newblock A personalized e-learning material recommender system.
\newblock In {\em International Conference on Information Technology and
  Applications}. Macquarie Scientific Publishing, 2004.

\bibitem{lu2010bizseeker}
Jie Lu, Qusai Shambour, Yisi Xu, Qing Lin, and Guangquan Zhang.
\newblock Bizseeker: a hybrid semantic recommendation system for personalized
  government-to-business e-services.
\newblock {\em Internet Research}, 20(3):342--365, 2010.

\bibitem{lu2015recommender}
Jie Lu, Dianshuang Wu, Mingsong Mao, Wei Wang, and Guangquan Zhang.
\newblock Recommender system application developments: a survey.
\newblock {\em Decision Support Systems}, 74:12--32, 2015.

\bibitem{lucas2013hybrid}
Joel~P Lucas, Nuno Luz, Mar{\'\i}A~N Moreno, Ricardo Anacleto, Ana~Almeida
  Figueiredo, and Constantino Martins.
\newblock A hybrid recommendation approach for a tourism system.
\newblock {\em Expert Systems with Applications}, 40(9):3532--3550, 2013.

\bibitem{mayer2015identifying}
Julia~M Mayer, Quentin Jones, and Starr~Roxanne Hiltz.
\newblock Identifying opportunities for valuable encounters: Toward
  context-aware social matching systems.
\newblock {\em ACM Transactions on Information Systems (TOIS)}, 34(1):1, 2015.

\bibitem{meza2019fuzzy}
Jaime Meza, Luis Ter{\'a}n, and Martha Tomal{\'a}.
\newblock A fuzzy-based discounts recommender system for public tax payment.
\newblock In {\em Applying Fuzzy Logic for the Digital Economy and Society},
  pages 47--72. Springer, 2019.

\bibitem{moreno2013sigtur}
Antonio Moreno, Aida Valls, David Isern, Lucas Marin, and Joan Borr{\`a}s.
\newblock Sigtur/e-destination: ontology-based personalized recommendation of
  tourism and leisure activities.
\newblock {\em Engineering Applications of Artificial Intelligence},
  26(1):633--651, 2013.

\bibitem{mustafa2015performance}
Ghulam Mustafa and Ingo Frommholz.
\newblock Performance comparison of top n recommendation algorithms.
\newblock In {\em 2015 Fourth International Conference on Future Generation
  Communication Technology (FGCT)}, pages 1--6. IEEE, 2015.

\bibitem{navimipour2014resource}
Nima~Jafari Navimipour, Amir~Masoud Rahmani, Ahmad~Habibizad Navin, and Mehdi
  Hosseinzadeh.
\newblock Resource discovery mechanisms in grid systems: A survey.
\newblock {\em Journal of Network and Computer Applications}, 41:389--410,
  2014.

\bibitem{nguyen2013web}
Thi Thanh~Sang Nguyen, Hai~Yan Lu, and Jie Lu.
\newblock Web-page recommendation based on web usage and domain knowledge.
\newblock {\em IEEE Transactions on Knowledge and Data Engineering},
  26(10):2574--2587, 2013.

\bibitem{orellana2015mining}
Claudia Orellana-Rodriguez, Ernesto Diaz-Aviles, and Wolfgang Nejdl.
\newblock Mining affective context in short films for emotion-aware
  recommendation.
\newblock In {\em Proceedings of the 26th ACM Conference on Hypertext \& Social
  Media}, pages 185--194. ACM, 2015.

\bibitem{park2017rectime}
Yoojin Park, Jinoh Oh, and Hwanjo Yu.
\newblock Rectime: Real-time recommender system for online broadcasting.
\newblock {\em Information Sciences}, 409:1--16, 2017.

\bibitem{pazzani2007content}
Michael~J Pazzani and Daniel Billsus.
\newblock Content-based recommendation systems.
\newblock In {\em The adaptive web}, pages 325--341. Springer, 2007.

\bibitem{polatidis2016multi}
Nikolaos Polatidis and Christos~K Georgiadis.
\newblock A multi-level collaborative filtering method that improves
  recommendations.
\newblock {\em Expert Systems with Applications}, 48:100--110, 2016.

\bibitem{porcel2010dealing}
Carlos Porcel and Enrique Herrera-Viedma.
\newblock Dealing with incomplete information in a fuzzy linguistic recommender
  system to disseminate information in university digital libraries.
\newblock {\em Knowledge-Based Systems}, 23(1):32--39, 2010.

\bibitem{qian2019ears}
Yongfeng Qian, Yin Zhang, Xiao Ma, Han Yu, and Limei Peng.
\newblock Ears: Emotion-aware recommender system based on hybrid information
  fusion.
\newblock {\em Information Fusion}, 46:141--146, 2019.

\bibitem{rodrigues2016efficient}
Celine~Michael Rodrigues, Sheetal Rathi, and Ganesh Patil.
\newblock An efficient system using item \& user-based cf techniques to improve
  recommendation.
\newblock In {\em 2016 2nd International Conference on Next Generation
  Computing Technologies (NGCT)}, pages 569--574. IEEE, 2016.

\bibitem{ruotsalo2013smartmuseum}
Tuukka Ruotsalo, Krister Haav, Antony Stoyanov, Sylvain Roche, Elena Fani,
  Romina Deliai, Eetu M{\"a}kel{\"a}, Tomi Kauppinen, and Eero Hyv{\"o}nen.
\newblock Smartmuseum: A mobile recommender system for the web of data.
\newblock {\em Web semantics: Science, services and agents on the world wide
  web}, 20:50--67, 2013.

\bibitem{salter2006cinemascreen}
James Salter and Nick Antonopoulos.
\newblock Cinemascreen recommender agent: combining collaborative and
  content-based filtering.
\newblock {\em IEEE Intelligent Systems}, 21(1):35--41, 2006.

\bibitem{santos2014extending}
Olga~C Santos, Jesus~G Boticario, and Diana P{\'e}rez-Mar{\'\i}n.
\newblock Extending web-based educational systems with personalised support
  through user centred designed recommendations along the e-learning life
  cycle.
\newblock {\em Science of Computer Programming}, 88:92--109, 2014.

\bibitem{shah2015hybrid}
Jaimeel~M Shah and Lokesh Sahu.
\newblock A hybrid based recommendation system based on clustering and
  association.
\newblock {\em Binary Journal of Data Mining \& Networking}, 5(1):36--40, 2015.

\bibitem{shambour2012trust}
Qusai Shambour and Jie Lu.
\newblock A trust-semantic fusion-based recommendation approach for e-business
  applications.
\newblock {\em Decision Support Systems}, 54(1):768--780, 2012.

\bibitem{tan2011using}
Shulong Tan, Jiajun Bu, Chun Chen, Bin Xu, Can Wang, and Xiaofei He.
\newblock Using rich social media information for music recommendation via
  hypergraph model.
\newblock {\em ACM Transactions on Multimedia Computing, Communications, and
  Applications (TOMM)}, 7(1):22, 2011.

\bibitem{tarus2017hybrid}
John~K Tarus, Zhendong Niu, and Abdallah Yousif.
\newblock A hybrid knowledge-based recommender system for e-learning based on
  ontology and sequential pattern mining.
\newblock {\em Future Generation Computer Systems}, 72:37--48, 2017.

\bibitem{vildjiounaite2009unobtrusive}
Elena Vildjiounaite, Vesa Kyll{\"o}nen, Tero Hannula, and Petteri Alahuhta.
\newblock Unobtrusive dynamic modelling of tv programme preferences in a
  finnish household.
\newblock {\em Multimedia systems}, 15(3):143--157, 2009.

\bibitem{wang2020government}
Yanan Wang, Airong Quan, Xiaonan Ma, and Junqing Qu.
\newblock E-government deep recommendation system based on user churn.
\newblock In {\em 2020 IEEE 8th International Conference on Smart City and
  Informatization (iSCI)}, pages 20--25. IEEE, 2020.

\bibitem{wu2014fuzzy}
Dianshuang Wu, Guangquan Zhang, and Jie Lu.
\newblock A fuzzy preference tree-based recommender system for personalized
  business-to-business e-services.
\newblock {\em IEEE Transactions on Fuzzy Systems}, 23(1):29--43, 2014.

\bibitem{wu2015item}
Hao Wu, Yijian Pei, Bo~Li, Zongzhan Kang, Xiaoxin Liu, and Hao Li.
\newblock Item recommendation in collaborative tagging systems via heuristic
  data fusion.
\newblock {\em Knowledge-Based Systems}, 75:124--140, 2015.

\bibitem{wu2019context}
Libing Wu, Cong Quan, Chenliang Li, Qian Wang, Bolong Zheng, and Xiangyang Luo.
\newblock A context-aware user-item representation learning for item
  recommendation.
\newblock {\em ACM Transactions on Information Systems (TOIS)}, 37(2):22, 2019.

\bibitem{yu2018pave}
Shuo Yu, Jiaying Liu, Zhuo Yang, Zhen Chen, Huizhen Jiang, Amr Tolba, and Feng
  Xia.
\newblock Pave: Personalized academic venue recommendation exploiting
  co-publication networks.
\newblock {\em Journal of Network and Computer Applications}, 104:38--47, 2018.

\bibitem{zhang2019personalized}
Jiang Zhang, Yufeng Wang, Zhiyuan Yuan, and Qun Jin.
\newblock Personalized real-time movie recommendation system: Practical
  prototype and evaluation.
\newblock {\em Tsinghua Science and Technology}, 25(2):180--191, 2019.

\bibitem{zhang2021dynamic}
Yanqing Zhang, Xuan Bi, Niansheng Tang, and Annie Qu.
\newblock Dynamic tensor recommender systems.
\newblock {\em Journal of Machine Learning Research}, 22(65):1--35, 2021.

\bibitem{zheng2011recommender}
Nan Zheng and Qiudan Li.
\newblock A recommender system based on tag and time information for social
  tagging systems.
\newblock {\em Expert Systems with Applications}, 38(4):4575--4587, 2011.

\bibitem{zhu2009personalized}
XiaoMing Zhu, HongWu Ye, and SongJie Gong.
\newblock A personalized recommendation system combining case-based reasoning
  and user-based collaborative filtering.
\newblock In {\em 2009 Chinese Control and Decision Conference}, pages
  4026--4028. IEEE, 2009.

\end{thebibliography}
\bibliographystyle{plain}

%\bibliographystyle{unsrt}
%\bibliography{oo}

\end{document}